\documentclass[a4paper,fleqn]{cas-sc}
\usepackage[authoryear]{natbib}
\usepackage{xcolor}
\usepackage{subcaption,mwe}
\usepackage{graphicx}

\def\tsc#1{\csdef{#1}{\textsc{\lowercase{#1}}\xspace}}
\tsc{WGM}
\tsc{QE}

\usepackage{lineno}

\begin{document}
\let\WriteBookmarks\relax
\def\floatpagepagefraction{1}
\def\textpagefraction{.001}

\shorttitle{Microscopic Propagator Imaging}    

\shortauthors{Zajac, Menegaz, \& Pizzolato}  

\title [mode = title]{Microscopic Propagator Imaging (MPI) with Diffusion MRI}
\author[1]{Tommaso Zajac}
\credit{Software, Formal analysis, Investigation, Validation, Visualization, Writing -- Original Draft}
\author[1]{Gloria Menegaz}[orcid=]
\credit{Investigation, Resources, Supervision, Funding acquisition, Writing -- Review \& Editing}
\author[2,3]{Marco Pizzolato}[orcid=0000-0003-2455-4596]
\cormark[1]
\ead{mapiz@dtu.dk pizzolato.marco.research@gmail.com}
\ead[url]{https://sites.google.com/site/pizzolatomarco/}
\credit{Conceptualization, Theory, Methodology, Software, Data curation, Formal analysis, Investigation, Supervision, Funding acquisition, Project administration, Writing -- Original Draft}

\affiliation[1]{organization={Department of Engineering for Innovation Medicine, University of Verona},
            city={Verona},
            country={Italy}}

\affiliation[2]{organization={Department of Applied Mathematics and Computer Science, Technical University of Denmark},
            city={Kgs. Lyngby},
            country={Denmark}}
\affiliation[3]{organization={Danish Research Centre for Magnetic Resonance, Department of Radiology and Nuclear Medicine, Copenhagen University Hospital - Amager and Hvidovre},
            city={Copenhagen},
            country={Denmark}}

\cortext[1]{Corresponding author}

\begin{abstract}
  We propose Microscopic Propagator Imaging (MPI) as a novel method to retrieve the indices of the microscopic propagator which is the probability density function of water displacements due to diffusion within the nervous tissue microstructures. Unlike the Ensemble Average Propagator indices or the Diffusion Tensor Imaging metrics, MPI indices are independent from the mesoscopic organization of the tissue such as the presence of multiple axonal bundle directions and orientation dispersion. As a consequence, MPI indices are more specific to the volumes, sizes, and types of microstructures, like axons and cells, that are present in the tissue. Thus, changes in MPI indices can be more directly linked to alterations in the presence and integrity of microstructures themselves. The methodology behind MPI is rooted on zonal modeling of spherical harmonics, signal simulation, and machine learning regression, and is demonstrated on both synthetic and Human Diffusion MRI data.
\end{abstract}


\maketitle

\section{Introduction}
One of the most prominent ways to harness the information obtained with Diffusion Magnetic Resonance Imaging (dMRI) is the calculation, for each voxel in the acquired image volume, of the Ensemble Average Propagator (EAP) \citep{callaghan1988nmr}. This encodes probabilistic information about diffusion of water particles (spins) within the tissue contained in the voxel's volume. Due to the interaction of water with the boundaries and membranes of axons, cells, and other microstructures, such probabilistic information speaks of the geometrical properties (like volume occupancies, sizes, and shapes) of the microstructures themselves.
A variety of methods have been developed to retrieve the EAP and its scalar descriptors, sometimes referred to as \emph{indices}. Among the most used ones we find the \emph{return-to-origin/axis/plane-probabilities} (RTOP, RTAP, RTPP), and the \emph{non-Gaussianity} (NG), which measures how different the EAP is from a Gaussian distribution.
The indices enable us to readily understand and compare differences between EAPs, and to characterize the diffusion process in different brain regions. For instance, RTOP is related to the reciprocal of the more common Diffusion Tensor Imaging (DTI) index called "mean diffusivity" (MD). When MD is high -- for instance in the brain ventricles where cerebrospinal fluid (CSF) diffusion is not hindered by barriers -- RTOP is low, and vice versa.
Unlike biophysical models, which interpret the dMRI signal a sum of contributions from a predefined set of water compartments (with or without exchange of water between them), EAP methods strives to keep a "compartment-agnostic" approach, thus virtually making no such assumptions. 
Technically, the EAP retrieved from dMRI describes the average displacement of the water molecules trajectory's center of mass \citep{mitra1995effects} accrued during the first and second gradient pulses of a pulsed gradient spin echo (PGSE) sequence \citep{stejskal1965spin}, at a given diffusion time. As such, it is affected by the mesoscopic organization of the tissue and in particular by the plurality of directions along which anisotropic structures, like axons, are aligned. In fact, even though the displacements inside of an axon mainly occur along its longitudinal direction (being diffusion restricted perpendicularly to it), the displacements inside of a multitude of axons occur in principle along as many directions as those axons are aligned. The EAP is thus clearly affected by the probability distribution of orientations of the anisotropic microstructures, known as the (fiber) \emph{orientation distribution function} (ODF), inside the voxel.
At the same time, however, the displacements are also determined by the geometry of axons, for instance by their sizes, but also by the relative abundance of axons themselves, cells (somas), and extra-axonal/cellular space.
As a consequence, the values of the scalar indices (like RTOP, RTAP, and RTPP) derived from the EAP may be regarded as ambiguous because they can be determined, at the same time, by the variety of orientations -- \emph{orientation dispersion} -- and by the presence/size/type of the different microstructures. Therefore, it is impossible to speculate on whether, in a specific voxel, an abnormally low RTAP (for example) is due to an atypical configuration of axonal orientations or to microstructural alterations in the presence/size/integrity of axons and cells.

We propose Microscopic Propagator Imaging (MPI) as a novel method to retrieve the propagator-based indices that are unaffected by the presence of multiple bundle directions and orientation dispersion.
This is achieved by isolating the indices of the microscopic propagator, which is the EAP of the rotational signal kernel that, when convolved with the ODF, "reconstructs" the dMRI signal. MPI thus isolates the contribution of the kernel signal to the overall EAP from that of the ODF. In other words, MPI indices only reflect the diffusion process pertaining to "non-dispersed" microstructures, and are only indicative of the abundance, geometries, sizes, thereby the integrity, of the microstructures present in the tissue.

\section{Theory}

The diffusion-weighted MRI signal arising from tissue microstructures can be expressed as a sum of the contributions from $N$ compartments. In particular, for a given b-value $b=q^2(\Delta-\delta/3)$ with $q=\gamma G\delta$ and $\Delta$, $\delta$, $G$ and $\gamma$ being respectively the gradient pulse separation, duration, amplitude, and the gyromagnetic ratio of the hydrogen atom, we have 

\begin{equation}
    S(b,\theta,\phi) = \sum_{n=1}^N S_{0n} \left\{ \sum_{l=0,even}^{L} \kappa_l^n(b|\textbf{p}_n)  \sum_{m=-l}^{l} c^{'}_{lm} Y_l^m(\theta,\phi)      \right\} 
    \label{eq_sum_signal_1}
\end{equation}
where $\theta$ and $\phi$ are the spherical coordinates of the normalized diffusion gradient vector (the sampled diffusion direction), $S_{0n}= S_n(0,\theta,\phi)= S_n(0)$ the constant that accounts for the relaxation and the volume fraction of the $n$-th compartment, and where $\textbf{p}_n$ is a generic vector of parameters for the $n$-th compartment. The spherical harmonics $Y_l^m(\theta,\phi)$ are those defined by \citet{descoteaux2007regularized}, and the $l$-degree subscript only assumes even values, corresponding to even spherical harmonics, because of the symmetry of the diffusion process.
Note that eq.~\ref{eq_sum_signal_1}, as it is usual in dMRI \citep{christiaens2020need},  implies that the compartments share the same ODF
\begin{equation}
    \textrm{ODF}(\theta,\phi) = \sum_{l=0,even}^{L} \sum_{m=-l}^{l} c^{'}_{lm} Y_l^m(\theta,\phi)  \ge 0  \, \forall \, \theta \in [0,\pi], \phi \in[0,2\pi]
    \label{eq_fodf}
\end{equation}
where $c^{'}_{lm}$ are the coefficients of the SH expansion of the ODF, and their values are typically unknown.
By rearranging the order of the sums 
\begin{equation}
    S(b,\theta,\phi) =   \sum_{l=0,even}^{L} \underbrace{\left\{ \sum_{n=1}^N S_{0n}\kappa_l^n(b|\textbf{p}_n)\right\}}_{K_l(b|\textbf{p})} \sum_{m=-l}^{l} c^{'}_{lm} Y_l^m(\theta,\phi)  
    \label{eq_sum_signal_2}
\end{equation}
the $N$ \emph{zonal functions} of zonal degree $l$, $\kappa_l^n(b|\textbf{p}_n)$ can be aggregated to form the more complex zonal functions $K_l(b|\textbf{p})$, where $\textbf{p}$ indicates a generic vector comprising all possible parameters. Note that these may generally include also parameters related to exchange between compartments. These new zonal functions are the zonal harmonic coefficients of the convolutional/rotational kernel signal, which may be reconstructed as
\begin{equation}
    S_K(b,\theta,\phi) =   \sum_{l=0,even}^{L}K_l(b|\textbf{p})Y_l^0(\theta,\phi)  
    \label{eq_kernel}
\end{equation}
up to a maximum SH degree $L$.
Further rearranging the summations, we observe that the zonal functions are actually part of the coefficients $c_{lm}(b)$ of a generic SH expansion of the directional dMRI on the shell at b-value $b$
\begin{equation}
\begin{split}
   S(b,\theta,\phi) &= \sum_{l=0,even}^{L}  \sum_{m=-l}^{l} \underbrace{ c^{'}_{lm} K_l\left(b|\textbf{p}\right)}_{c_{lm}(b)} Y_l^m(\theta,\phi)\\    
   &= \sum_{l=0,even}^{L} \sum_{m=-l}^{l} c_{lm}(b) Y_{l}^{m}(\theta,\phi)
\end{split}
\label{eq_sum_signal}
\end{equation}
from which the coefficients can be estimated by solving a linear system.
At a given b-value, we can then calculate the $l$-band power spectra for each zonal degree $l=0,2,\dots,L$ as
\begin{equation}
\begin{split}
        \Theta_l(b) &= \sum_{m=-l}^{l} c_{lm}(b)^2\\
        &= K_l\left(b|\textbf{p}\right)^2\sum_{m=-l}^{l} \left(c^{'}_{lm} \right)^2\\
\end{split}
    \label{eq_lband_spectrum}
\end{equation}
which are rotational invariants, meaning that no matter how the dMRI signal is rotated, the power spectra will remain constant \citep{zucchelli2020computational}.
Of more interest are the ratios between spectra of the same degree $l$ of dMRI signals with different b-values, as they additionally are independent from the ODF \citep{reisert2017disentangling,pizzolato2023axial}.
In particular, for two non-zero b-values $b_i \neq b_j$ we can define the root-square $l$-band power spectral ratios (which we will simply call \emph{ratios}) as
\begin{equation}
\alpha_l\left(b_i,b_j|\textbf{p}\right) := \sqrt{\frac{\Theta_l(b_i)}{\Theta_l(b_j)}} = \left| \frac{K_l\left(b_i|\textbf{p}\right)}{ K_l\left(b_j|\textbf{p}\right)} \right|
\label{eq_ratios}
\end{equation}
which clearly only depend on the parameter vector $\textbf{p}$, the b-values, and the functional forms of the zonal functions.
Given any two signal shells, both expanded to a maximum degree $L$ it is then possible to obtain $L/2+1$ independent ratios.
\citet{reisert2017disentangling} regressed similar ratios with respect to the parameters of a signal kernel corresponding to the "standard model" of white matter diffusion \cite{novikov2019quantifying}, discovering that degeneracy \citep{jelescu2016degeneracy} occurred when using PGSE data.
Therefore, not all parameters of the model (fractions and diffusivities) could be uniquely identified.
Here we make the hypothesis that, instead, the indices of the kernel's propagator, in particular indices like the "return-to probabilities" and non-Gaussianity, can be recovered from the ratios.
If so, we would be able to train a machine learning regression method that learns the mapping between the ratios $\alpha_l(b_i,b_j|\textbf{p})$ (plus eventually other signal rotational invariants) and the indices of the kernel's propagator, which are the MPI indices of the \emph{microscopic propagator}.

\section{Methods}
The methodology consists of a \emph{learning} part followed by an \emph{inference} part. The learning part is focused on training the parameters of a regressor that maps the indices of the microscopic propagator, \emph{labels}, to a mix of rotationally invariant and ODF-independent (the ratios) \emph{features} of the multi-shell dMRI signal. The training is performed using synthetically generated labels and features (matching the protocol of the acquired dataset). After the training has been completed, the features of the acquired signal are calculated and fed as input to the regressor to obtain the estimates of the MPI indices.  
\begin{figure}[h!]
    \centering
    \includegraphics[width=1\textwidth]{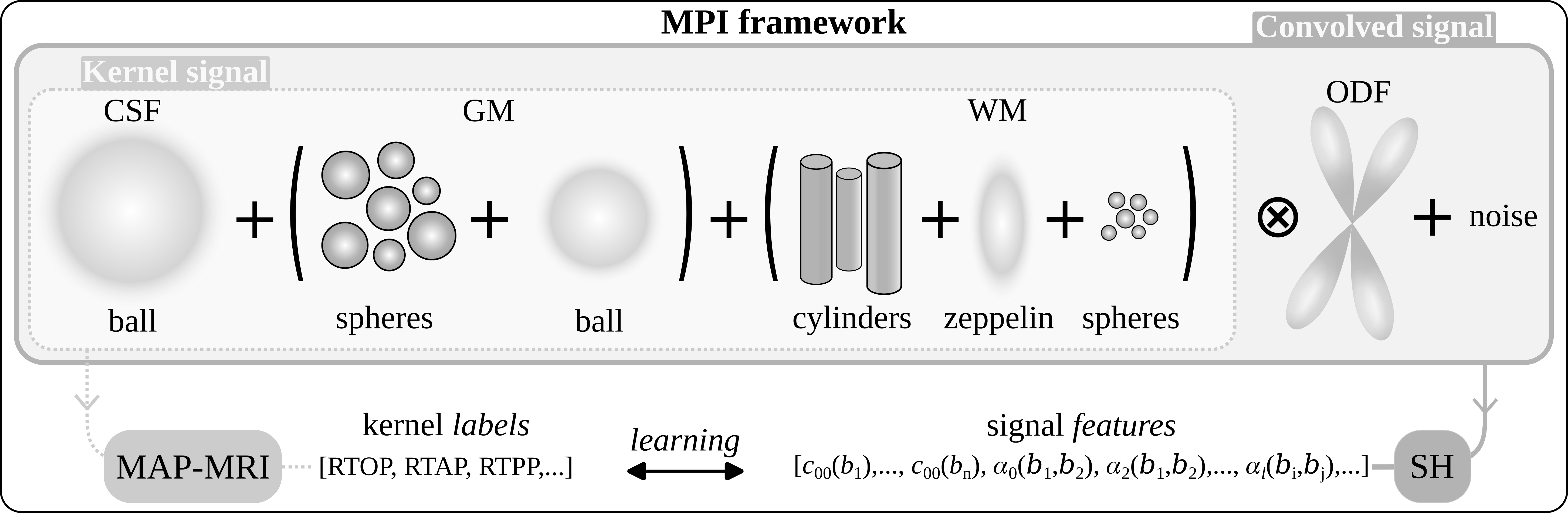}
    \caption{The framework implementing the concept of microscopic propagator imaging for what concerns the generation of (signal) features and (kernel) labels to be learned.}
    \label{fig_mpi_framework}
\end{figure}
\subsection{Microscopic propagator indices}
To perform the regression between the ratios and the MPI indices it is necessary to generate a database of synthetic signals on which to compute the ratios (\emph{features}) and the sought indices of the microscopic propagator such as $\textrm{RTOP}_{\textrm{MPI}}$, $\textrm{RTAP}_{\textrm{MPI}}$, etc. (\emph{labels}). The overall methodology is illustrated in Fig.~\ref{fig_mpi_framework}.
The generation of the synthetic signal features needs to target the desired acquisition scheme, namely the set of gradient directions $(\theta,\phi)$, b-values ($b$), pulse duration ($\delta$), pulse separation ($\Delta$), and optionally the echo time ($\textrm{TE}$).
In this work, we simulated the signal with a multi-compartment model comprising three main tissues: white matter (WM), gray matter (GM), and cerebrospinal fluid (CSF). 
CSF was modeled as isotropic Gaussian diffusion (ball) with intrinsic diffusion coefficient in range $[2.9\text{e\;-}9,3.1\text{e\;-}9]$m$^2$/s.
GM was modeled as extra-cellular diffusion using a ball of diffusivity in range $[0.85\text{e\;-}9,1.3\text{e\;-}9]$m$^2$/s, plus an intra-cellular part, representing neuronal and other cellular somas, comprising of an ensemble of spheres which radii follow a Gamma distribution with mode in range $[3,30]\mu{\textrm{m}}$, using intrinsic diffusivity of $1.7\text{e\;-}9$ m$^2$/s \citep{fick2019dmipy}.
White Matter (WM) was modeled as the sum of three compartments. An intra-axonal part comprising of cylinders with radii following a Gamma distribution with mode in range
 $[0.2, 0.25]\mu{\textrm{m}}$, using intrinsic diffusivity in range $[1.5\text{e\;-}9, 3.0\text{e\;-}9]$m$^2$/s 
 \citep{pizzolato2023axial}.
 A cellular component comprising of a set of smaller spheres was also added to account for the presence of glial and endothelial cells. Finally, an extra-axonal part was added and modeled with a zeppelin (axisymmetric tensor) of axial and radial diffusivities in range $[0.9\text{e\;-}9, 1.6\text{e\;-}9]$m$^2$/s.
For the attenuation due to spheres and cylinders the Gaussian Phase Approximation formulas were used \citep{neuman1974spin,balinov1993nmr,vangelderen1994evaluation}.
The Gamma distributions for the sphere and cylinder radii were defined by the two shape parameters to obtain the desired mode.
In all cases, size distributions were based on informed guesses \citep{Liewald2014} (and wide ranges) to cover all possible biological variability.
Transverse relaxation attenuation was also simulated for each compartment, to improve the realism of the simulated signal, to correctly account for noise, and to account for potential differences in the echo time of the targeted acquisition scheme.
The ranges of $T_2$ times were in [0.035, 0.065]s for the intra-axonal compartment, [0.06,0.08]s for the extra-axonal one, [0.02, 0.04]s for the spheres, [0.07, 0.1]s for the extra-axonal/cellular compartments, while CSF's $T_2$ was fixed to 2s.
The tissue fractions and the relative fractions within the tissue were randomly generated following Dirichlet or uniform distributions. By randomly generating 2 million different realizations of the modeling parameters, we obtained as many kernel signals through the evaluation of the resulting multi-compartment model, which was aligned along a specific direction (e.g. z-axis).
Then each kernel signal was fitted using the MAPL \citep{fick2016mapl} version of MAP-MRI \citep{ozarslan2013mean} to obtain microscopic MPI indices. In addition to the mentioned $\textrm{RTOP}$, $\textrm{RTAP}$, and $\textrm{RTPP}$, parallel and perpendicular non-Gaussianity indices NG$_\parallel$, NG$_\perp$ were also estimated. These constitute the ground-truth labels to be learned.

\subsection{Signal features}
The kernel signal was then convolved by an ODF selected from a database of realistic ODFs obtained from Multi-shell Multi-tissue constrained spherical deconvolution \citep{jeurissen2014multi} (MSMT-CSD) on a variety of datasets.
We made sure to convolve with ODFs with spherical variance significantly higher than zero (we selected ODFs with variance above the 80th percentile in our database) to avoid causing the ratios of degree $l\ge 2$ to be null \citep{pizzolato2022axonal}, which would prevent any effective learning during the regression phase, however a different approach may be envisioned.
Noise was added to the convolved signal according to a specified signal-to-noise ratio (SNR), following a Gaussian noise distribution. This was chosen since we preprocessed the actual dMRI data using a Variance Stabilization Transformation (VST) that removes the Rician bias \citep{foi2011noise}.
The noise was randomly selected for each signal example with standard deviation extracted from a uniform distribution with lower and upper bounds optimized to find those that yielded the best performance on the testing datasets.
The noisy signal was divided by the simulated non-attenuated signal ($b=0$) to which noise was also added and averaged over the number of $b=0$ volumes in the acquisition scheme to replicate what was done for the acquired data.
The obtained noisy attenuation signals were used to obtain the SH coefficients \citep{descoteaux2007regularized} for each shell $\{c_{lm}(b)\}$. With these, we calculated the power spectra of each shell and, from these, the ratios between them for the b-value pairs (1000, 3000), (1000, 5000), (1000, 10000), (3000,5000), (3000,10000), (5000, 10000).
This process enables us to provide examples of spectra and ratios where the noise distribution is the most similar to that of the features calculated on the acquired data.
The SH coefficients obtained from the noisy attenuation signals were up to SH order 12, or 6 depending on the number of directions available for each shell.
Therefore, for each shell, up to 4 ($\ell=0,2,4,6$)  or 7 $\ell$-band power spectra ($\ell=0,2,4,6,8,10,12$) were obtained. The final set of features included, in addition to the ratios, the so-called DC component, $c_{00}(b)$, of each shell, which corresponds to the spherical mean (powder averaged) attenuations. It is important to include these, since they enable the regressor to discriminate between the presence and absence of signal (in addition to providing other significant information). In fact, the quotient resulting from the ration of two terms (the power spectra) will be dominated by noise, thus uninformative, in the case when the terms are close to zero (for instance, in the brain ventricles).

\subsection{Machine learning regression}
\begin{figure}[h]
    \centering
    \includegraphics[scale=0.5]{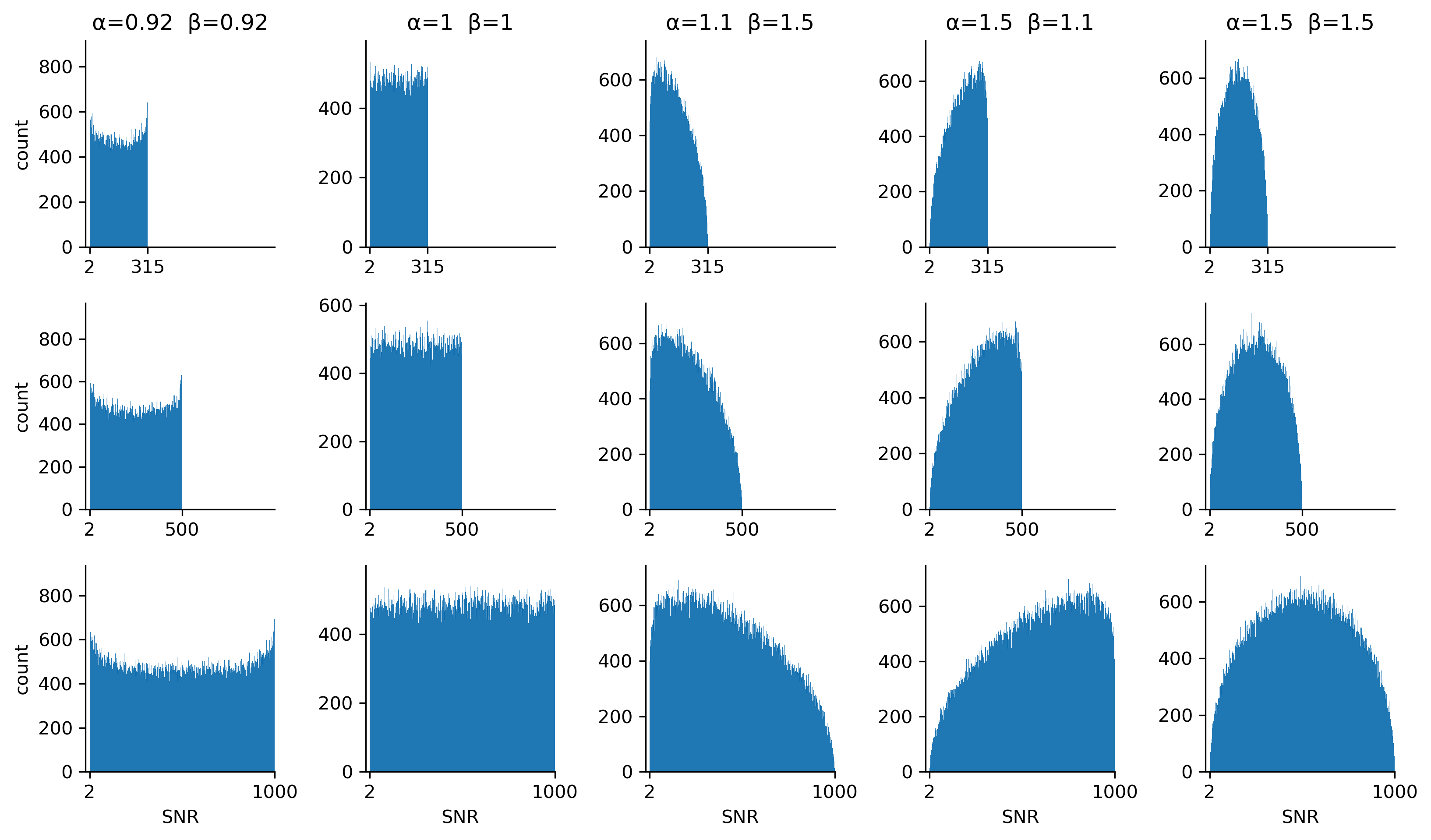}
    \caption{Probed SNR distributions for adding noise on the training data.}
    \label{fig_snr_dist}
\end{figure}

The mapping between the 2 million sets of features and the MPI indices was performed with a random forest (RF) regressor \citep{breiman2001random}. Each decision tree in the forest was trained on a random subset of the training set (bootstrap aggregation) to improve performance and reduce the risk of overfitting. The goodness of a node split was determined by the Friedman mean squared error \citep{Friedman2001}. This was found to give a lower out-of-bag error compared to the conventional means squared error, which is connected to the fact that the selected features are correlated. Regression was performed for fifteen different training datasets that differed from each other in the distribution used to randomly sample the SNR. The SNR distributions were obtained by varying the SNR ranges ($[2,315], $$[2,500]$ and $[2,1000]$) and the shape parameters ($\alpha$ and $\beta$) of a Beta distribution as (0.92,0.92), (1,1), (1.1,1.5), (1.5,1.1) and (1.5,1.5), as shown in \ref{fig_snr_dist}. The first SNR range was extracted from real data (assuming a $\sqrt{2}$ increase in SNR due to denoising), the second and third were investigated to study the effect of a wider range of SNR in the training data. The shape parameters of the Beta distribution were arbitrary chosen to have a uniform SNR distribution (1,1), to give more weight to the range's extreme (0.92,0.92), or to weight more signal with low (1.1,1.5), high (1.5,1.1) or mid (1.5,1.5) SNR in the training set.
The best SNR distribution for generating the synthetic data used to train the model was chosen as that minimizing the absolute error between the predicted and ground-truth labels in the testing set (unseen instances) and the Wasserstein distance \citep{villani2009wasserstein} between the distribution of the predicted labels and the distribution of the ground-truth ones in the testing set. The best performance was found using the Beta distribution with $\alpha=\beta=1.5$ in the range $[2,1000]$.
Once the SNR distribution was established, the hyperparameters of the RF regressor were tuned with a 3-fold cross-validation grid search based on the mean squared error (MSE) and the coefficient of determination ($R^2$). The hyper-parameters included in the search were the maximum depth of the trees ([automatic, 10, 20]), the maximum number of features to take into consideration when splitting a node ([100\%, 80\%, 'sqrt'] w.r.t. the number of features), the minimum number of samples in a node required to be considered a leaf ([2, 5]), the number of samples to draw with replacement from the training set to train a decision tree ([100\%, 75\%] w.r.t. the total number of samples in the training set), and the number of decision trees in the ensemble ([100, 200]). The best performance was found for a maximum depth of 20, a maximum number of features to consider for splitting a node according to the square root of the total number of features, a minimum of 5 samples in the leaf node, 0.75 of the total number of samples in the training set (1.5 million) to train each decision tree (leaving the rest for validation) and 100 decision trees. The resulting RF regressor was used for testing using unseen testing datasets.

\subsection{Synthetic testing datasets}
We generated three different test datasets that included 1 million unseen instances each.
The datasets were designed to match the expected tissue proportions and SNR ranges of the WM, GM, and CSF tissues of an HCP dataset. The data generation model used was the same as for the training data.
For each tissue type, the SNR range was established based on that estimated in a real subject based on a segmentation of the WM, GM, and CSF regions obtained by estimating the corresponding tissue fractions with MSMT-CSD. Therefore, we simulated an SNR range of $[11,51]$ for WM, of $[16,77]$ for GM, and of $[24,152]$ for CSF, which stems from the assumption of a $\sqrt{2}$ increase in SNR (compared to the original data) due to denoising.
In addition, the simulated proportions of the volume fractions of WM, GM, and CSF
were simulated by modifying the concentration parameters of the Dirichlet distributions to mimic the fraction distributions found in the corresponding segmentation masks.
The random seed was different from that used to generate the training dataset.

\subsection{MRI data}
Data from 4 subjects of the Human Connectome Project (HCP) Adult Diffusion database collected at Massachusetts General Hospital (MGH) were used for the study \citep{setsompop2013pushing}.
The single encoding diffusion-weighted images (DWIs) were acquired using a monopolar PGSE sequence and are organized into 40 $b=0$ volumes plus four shells: 64 directions for $b=1000\,\textrm{s/mm}^2$ and for $b=3000\,\textrm{s/mm}^2$, 128 directions for $b=5000\,\textrm{s/mm}^2$ and 256 directions for $b=10000\,\textrm{s/mm}^2$.
Other parameters were: repetition time was TR$=8.8\textrm{s}$, echo time TE$=57\textrm{ms}$, gradient pulse duration $\delta=12.9\textrm{ms}$ and separation $\Delta=21.8\textrm{ms}$, field of view FOV$=210\textrm{mm}\,\textrm{x}\,210\textrm{mm}$, acquisition matrix of $140\,\textrm{x}\,140$ voxels, for 96 slices producing a $1.5\textrm{mm}$ isotropic voxel size, iPAT=3, and partial Fourier 6/8.
A complete description of the parameters is available on the Human Connectome website\footnote{\url{https://www.humanconnectome.org/study/hcp-young-adult/document/mgh-adult-diffusion-data-acquisition-details}}.
Data were acquired with a Siemens 3T MAGNETOM ConnectomA syngo MR D11 (Siemens Heathineers, Erlangen, Germany) capable of reaching a gradient strength of approximately 300$\textrm{mT/m}$.
Data were pre-processed as in \citet{pizzolato2023axial}, following a slightly modified version of the pipeline described by \citet{ma2020denoise}.
The denoising is based on local principal component analysis using optimal shrinkage with respect to the mean squared error \citep{gavish2017optimal} which follows a Rician variance stabilization transformation (VST) \citep{foi2011noise}, so that the denoising removes the Rician bias and increases the SNR of the images. Gibbs ringing removal according to \citet{kellner2016gibbs} was applied using the implementation available in MRtrix3\footnote{\url{https://www.mrtrix.org/}} \citep{tournier2019mrtrix3}, followed finally by FSL's eddy correction \citep{andersson2016integrated}.

\subsection{Implementation and visualization}
The code has been implemented in Python\footnote{\url{https://www.python.org/}}.
The visualization of the results was performed with Matplotlib\footnote{\url{https://matplotlib.org/}} \citep{Hunter_2007}.
The evaluation of the equations was carried out with NumPy\footnote{\url{https://numpy.org/}} \citep{2020NumPy-Array} and SciPy\footnote{\url{https://www.scipy.org/}} \citep{2020SciPy-NMeth},  the RF regressor model and the cross validation grid search were implemented with the Python library scikit-learn\footnote{\url{https://www.scikit-learn.org/}} \citep{scikit-learn}.
When constructing both the training and testing datasets, the SH coefficients were estimated using Laplace-Beltrami regularization \citep{descoteaux2007regularized} with weight $\lambda=1\text{e\;-}6$ and maximum order of either $L=6$ or $L=12$ (depending on the number of directions in the shell). To generate the datasets with a large number of samples, we took advantage of the parallelization of the computation with Ray\footnote{\url{https://www.ray.io/}} \citep{moritz2018ray}.
The MAPL model used to recover the ground-truth value of the microscopic indices was obtained from Dipy\footnote{\url{https://dipy.org/}} \citep{garyfallidis2014dipy} using a radial order of 12, anisotropic scaling and a Laplacian regularization weight of 0.01. To obtain MAP-MRI data on the four subjects, MAPL was used with radial order 8, anisotropic scaling, positivity constraint, and Laplacian weighting of 0.2, as these parameters enabled us to obtain maps with reduced number of "failing" voxels. The simulation of the kernel signals and convolved signals was carried out using Dmipy\footnote{\url{https://github.com/AthenaEPI/dmipy}} \citep{fick2019dmipy}.

\section{Results}
%
\begin{figure}[h!]
    \centering
    \includegraphics[width=1\textwidth]{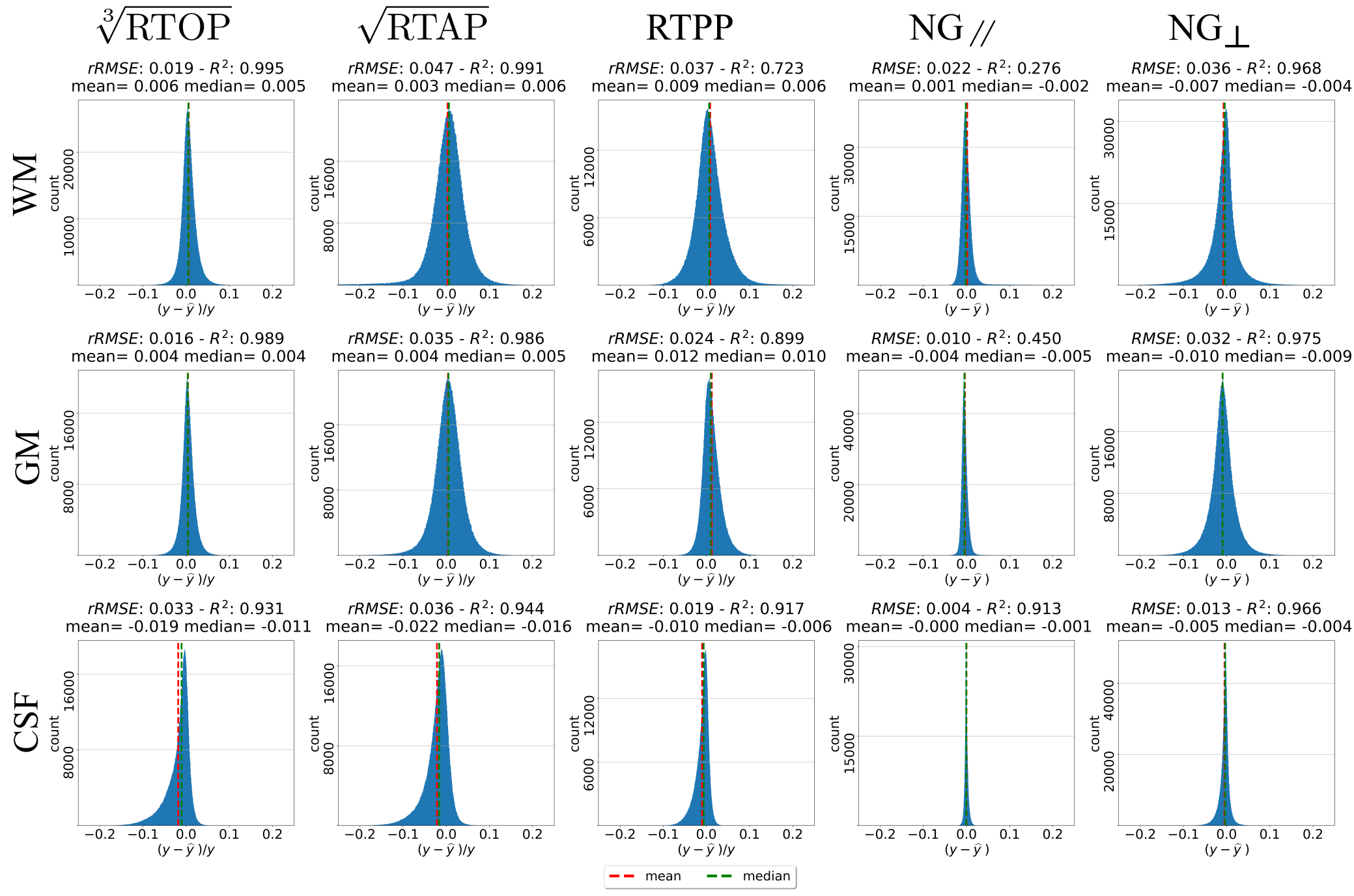}
    \caption{Error distributions and performance metrics across the white matter (WM), gray matter (GM), and cerebrospinal fluid (CSF) synthetic testing dataset.}
    \label{fig_simu_err}
\end{figure}
We compared the ground-truth MPI indices with those predicted, for each of the synthetic testing datasets of WM, GM and CSF. Fig.~\ref{fig_simu_err} presents the error distributions, along with their mean and median values, and key performance metrics such as the Coefficient of Determination ($R^2$) and Root Mean Squared Error (RMSE). RMSE was calculated for relative errors (rRMSE) for RTOP, RTAP, and RTPP, while non-relative errors were used for NG indices, since they fall within the range [0,1].
Most of the relative error distributions for microscopic RTOP, RTAP and RTPP fall mainly within -10$\%$ and 10$\%$, which, together with a $R^2$ above 0.7, indicates good regression performance. A similar consideration applies to NG$_{\perp}$ (where the error distribution is even narrower). However, a low $R^2$ for NG$_{\parallel}$ indicates poor learning performance in this case, with and improved $R^2$ for GM and CSF compared to WM.
\begin{figure}[h]
    \centering
    \includegraphics[width=1\textwidth]{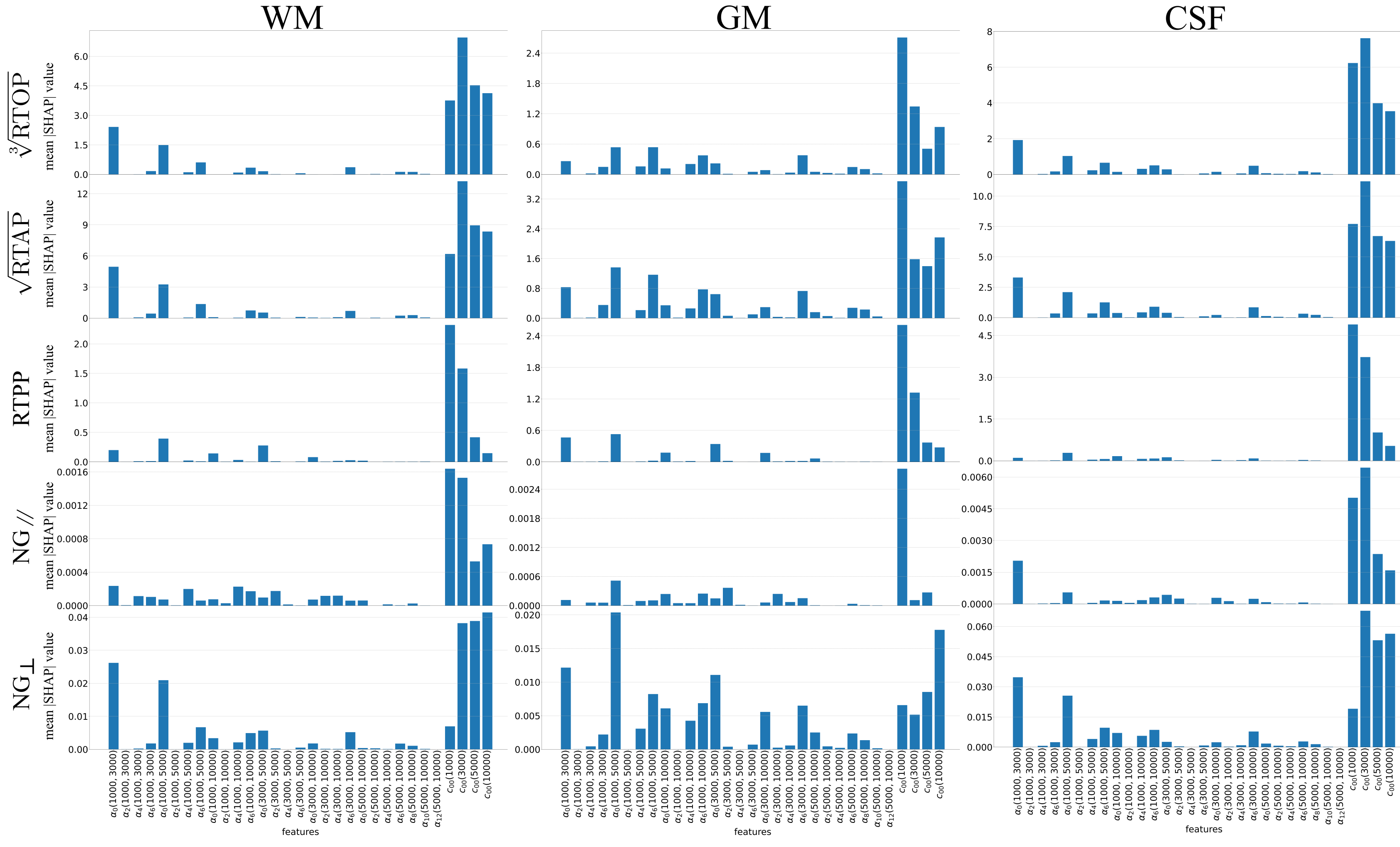}
    \caption{Mean absolute SHAP values representing the features importance for the selected trained RF regressor. Results were computed by randomly extracting 10000 samples from each of the WM, GM, and CSF testing datasets. A higher value indicates a higher importance of the feature.}
    \label{fig_shap}
\end{figure}

We calculated the SHapley Additive exPlanations (SHAP) values \citep{lundberg2017unified} using the tree explainer \citep{lundberg2020local} on the RF model to obtain a reliable quantification of the importance of the features. More conventional measures may be biased when features are correlated \citep{strobl2008conditional,nembrini2018revival}. Fig.~\ref{fig_shap} illustrates the importance of features calculated on a representative subsample (10000 examples) of each testing WM, GM, and CSF dataset. The DC components are the most important features. This is expected since the $c_{00}$ coefficients of the SH expansion are related to spherical mean of the shells and determine most of the kernel signal variability, hence of its propagator. Results demonstrate the importance of the ratios, which is modulated by the dataset, the considered MPI index, and the zonal degree. Notably, ratios have a particularly prominent importance in the determination of $\textrm{NG}_{\perp}$ and in the GM dataset.

\begin{figure}[h!]
    \centering
    \includegraphics[height=0.88\textheight]{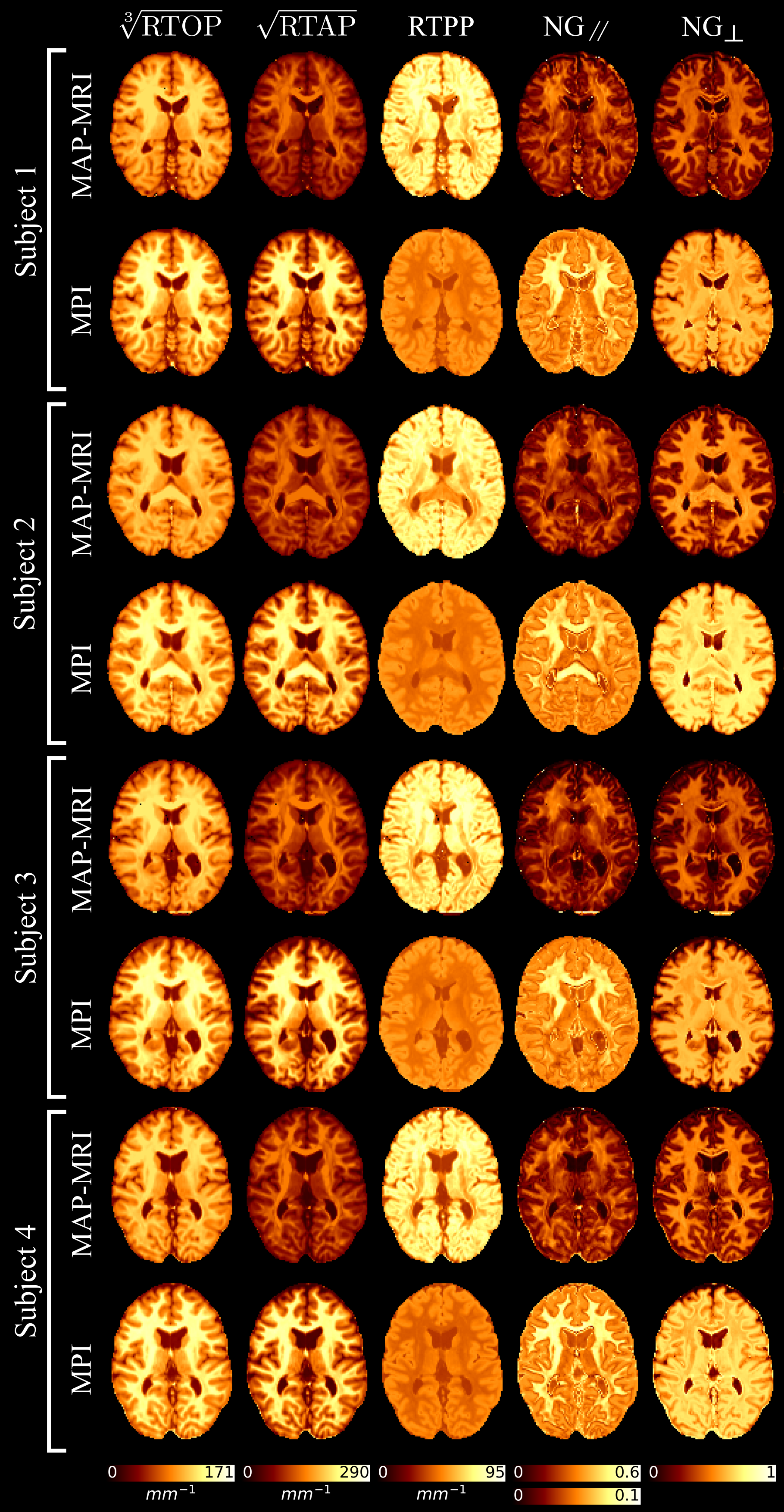}
    \caption{Maps of propagator indices obtained with MAPL implementation of MAP-MRI and with MPI for four subjects of the HCP dataset. MAPL's results were clipped between 0 and the 99$^{\text{th}}$ percentile to remove outliers. The color bars' upper limits were adjusted by considering the highest of values between MAPL and MPI. In the case of NG$_{\parallel}$ two different color bar ranges are used ([0,0.6] for MAPL and [0,0.1] for MPI) to adjust for the markedly different value ranges.
    }
    \label{fig_map}
\end{figure}
\begin{figure}[h]
    \centering
    \includegraphics[width=1\textwidth]{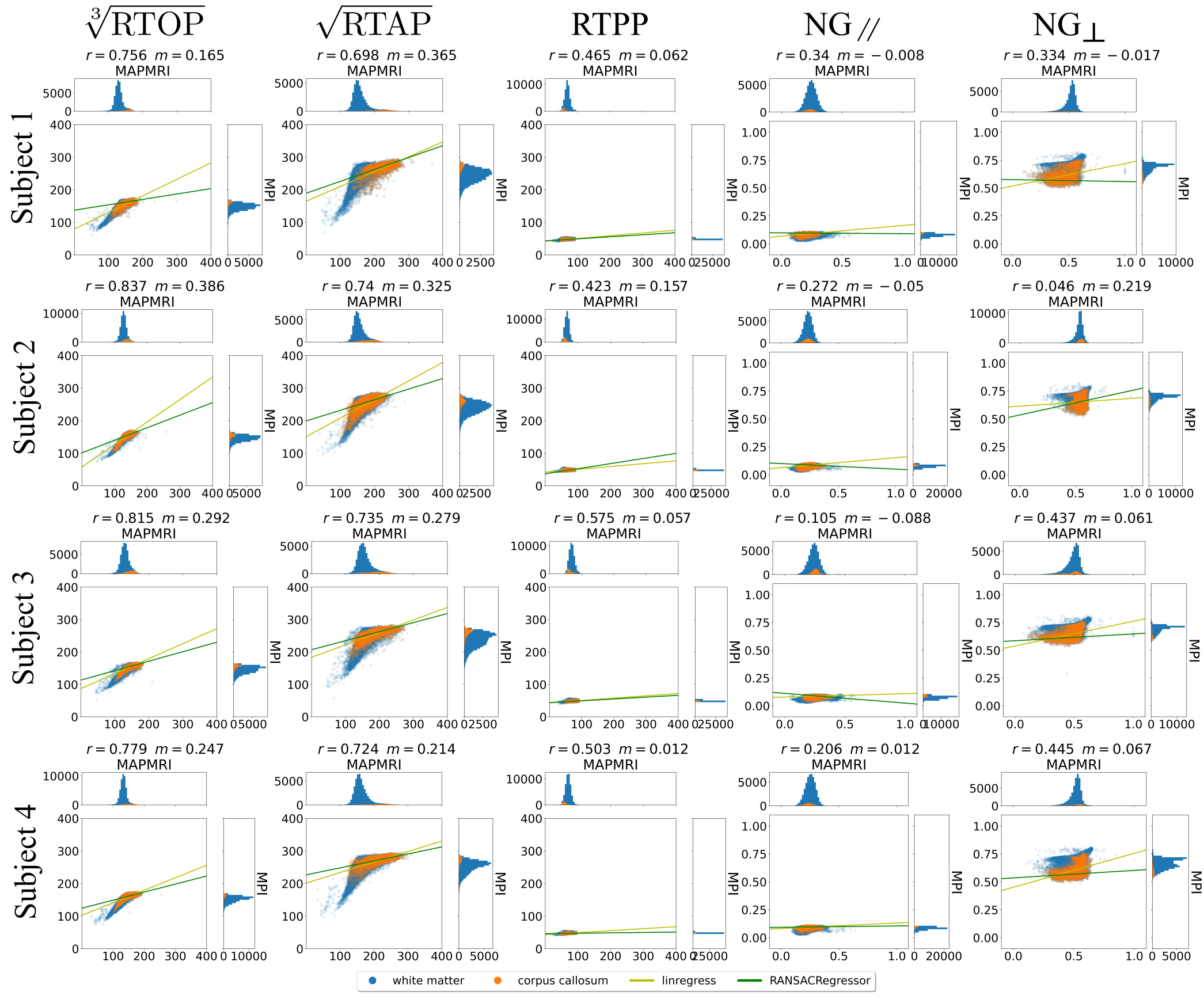}
    \caption{Correlation between MAP-MRI (MAPL) and MPI in real subjects ROIs. A white matter mask was created by selecting the voxels with a predominant fraction of white matter (the fractions of the tissues were recovered using MSMT-CSD) and a fractional anisotropy (FA) above 0.5. Additionally, for each subject, the corpus callosum mask was obtained with manual segmentation.}
    \label{fig_corr}
\end{figure}

MPI indices were compared with the corresponding MAP-MRI ones (processed with MAPL) on the four healthy subjects of the HCP dataset (Fig.~\ref{fig_map}).
Fig.~\ref{fig_corr} illustrates the correlation scatter plots of the MPI versus MAP-MRI indices for voxels located in white matter (blue) and for voxels located in the region comprising the corpus callosum (orange). For this region, the plots include the regression lines obtained with linear regression (yellow) and with the robust estimator RANSAC \citep{fischler1981random} (green). In addition, the slope of the RANSAC regression line ($m$) and the Pearson correlation coefficient ($r$) are reported.

The comparison between MPI indices and those obtained with MAP-MRI on the tested subjects (Fig.~\ref{fig_corr}) reveals a good correlation for RTOP and RTAP in regions, like the corpus callosum, where we expect the tissue to be composed of a single bundle of coherently aligned axons. In fact, there, the EAP and the microscopic propagator should share similar properties albeit the latter being additionally insensitive to within-bundle orientation dispersion, i.e. \emph{microdispersion}. 
MPI's insensitivity to microdispersion may explain the low correlation slope for RTPP. In fact, according to synthetic results shown in Fig.~\ref{fig_simu_err}, MPI's RTPP is robustly estimated therefore we can exclude the estimation error being responsible for the slope. Instead, we may focus on the effect of microdispersion on the EAP's RTPP as opposed to MPI. MPI's return-to-plane probability is approximately proportional to the inverse of the axial diffusivity of the axonal compartment (the most abundant) in white matter. While \emph{small} variations in MPI's RTPP may occur due to slightly changing microstructural properties, we see no reason to expect \emph{large} variations of it in (healthy) homogeneous regions like the corpus callosum.
We indeed observe that MPI's RTPP is more constant compared to that estimated with MAP-MRI (Fig.~\ref{fig_map}).
We then suggest that the variability of MAP-MRI's RTPP estimates may thus reflect the differences in microdispersion between the voxels in the ROIs.
To corroborate this observation, MPI's RTPP is lower than MAP-MRI's counterpart which confirms findings according to which, correspondingly, the axial axonal diffusivity obtained with a method that removes microdispersion is higher than what calculated otherwise \citep{pizzolato2023axial}.
A similar consideration can be made for the perpendicular non-Gaussianity, $\textrm{NG}_{\perp}$. In the presence of orientation dispersion, the estimated EAP-based $\textrm{NG}_{\perp}$ will have non-negligible contributions from the axonal diffusion along axons that are not perfectly aligned with the EAP's principal diffusion direction. This "washing" effect would then reduce the effective value of the EAP-based $\textrm{NG}_{\perp}$, since axons are expected to display a more marked non-Gaussianity perpendicularly to their axes compared to along the direction aligned with (parallel to) them. This phenomenon does not occur, or it does with a lesser extent, in the case of MPI where we observe larger $\textrm{NG}_{\perp}$ values (last column of Fig.~\ref{fig_map}).
Hence, it should not surprise that the correlation slopes of MPI versus MAP-MRI indices in Fig.~\ref{fig_corr} differ from the identity, as this is an expected outcome. MPI's parallel non-Gaussianity $\textrm{NG}_{\parallel}$ deserves special consideration. Maps in Fig.~\ref{fig_map}  suggest a relatively high value in white matter, which may indicate the presence of effects due to tortuous (or generally not cylindrical) axonal geometries and/or the presence of cellular somas. However, the amount of parallel non-Gaussianity is visibly lower compared to MAP-MRI (note that for MPI the color bar has a lower maximum extent). This is also expected, since part of MAP-MRI's $\textrm{NG}_{\parallel}$ can be explained, similarly to the other indices, by orientation dispersion. However, MPI's $\textrm{NG}_{\parallel}$ is the only index which has unsatisfactory $R^2$ values (see Fig.~\ref{fig_simu_err}), therefore the corresponding maps cannot be trusted.
A reason for the low performance of the parallel non-Gaussianity may be found in the training data for, other than the non-Gaussianity due to the presence of cells of different sizes, no sources of parallel non-Gaussianity for axons have been simulated. This corresponds to having a training dataset where the axonal contribution to $\textrm{NG}_{\parallel}$ is zero. As a consequence, the synthetic signal features of training examples resembling white matter signals (where axons are the main compartment) may likely be mapped to the value of the \emph{axonal} parallel non-Gaussianity, hence to zero.
This is not the case of the perpendicular direction (that of $\textrm{NG}_{\perp}$) where both the ensemble of cells with different radii, and the ensemble of cylinders of various diameters induce non-Gaussianity. 

\section{Discussion}
The sets of spherical harmonic coefficients estimated for each PGSE shell are naturally linked to the diffusion process occurring within the voxel's tissue microstructure. Under the condition that the signal may actually be obtained as the convolution between a signal rotational \emph{kernel} and an \emph{orientation distribution function} (ODF), which may be the main limiting assumption of MPI, the ratios between corresponding coefficients of any two shells -- as well as the ratios between corresponding $l$-band power spectra -- are \emph{solely} linked to the properties of the kernel itself. In fact, the ratios are \emph{independent} from the ODF. Although this property reveals useful in estimating axial and radial axonal diffusivities, and radii (even accounting for the presence of somas) when using data at strong diffusion weightings \citep{pizzolato2023axial}, the ratios do not prevent degeneracy when attempting to estimate the parameters of the relatively simple "standard" model of (white matter) diffusion \citep{reisert2017disentangling} using a multi-shell scheme (even as rich as the HCP data used here).
However, the results in this work demonstrate that the ratios are directly sensitive to the indices of the kernel's (microscopic) propagator, when collecting data over multiple shells with diffusion weightings defined in a "q-space sense" (where the diffusion time $\Delta-\delta/3$ and the pulse duration $\delta$ are kept constant).
This fact opens up the new possibility of performing \emph{microscopic propagator imaging} (MPI).

The majority of microscopic propagator indices were robustly estimated. However, we have observed that parallel non-Gaussianity $\textrm{NG}_{\parallel}$ is not reliable since it had a low regression performance.
We cannot say at this moment whether this is due to an intrinsic low sensitivity of the selected features to that specific index or if this is due to the parallel non-Gaussianity not being simulated for axons in the training dataset, as mentioned previously.
Nevertheless, this consideration leads us to believe that more sophisticated synthetic signal generation strategies, for example via Monte Carlo simulation, would improve the realism of the MPI indices and likely enable the learning of other features of the microscopic propagator. It is worth mentioning that the presented framework can in principle accommodate the presence of exchange between tissue compartments in the simulated kernel signal, although a quantification of it may not be straightforward or even possible with the current $q$-space acquisition scheme.
A more realistic simulation could also benefit some technical choices in the implementation of MPI. For example, using Monte Carlo simulation, it would be possible to completely bypass the step of fitting MAP-MRI to the simulated kernel signal, since RTOP, RTPP and the other indices may be calculated directly from the trajectories of the simulated random walkers.
This would additionally provide the benefit of avoiding the tuning of parameters for obtaining the microscopic propagator indices, such as the maximum basis order of MAP-MRI and MAPL's Laplacian regularization.
In our simulations, these parameters had an effect on the fitting quality, which also depended on the degree of isotropy, or anisotropy, of the simulated kernel signal. Hence, our choices were based on a trade-off between being able to reconstruct very anisotropic kernels and not having too much overfitting in the case of kernels that were mostly isotropic. In practice, after visual inspection, we adopted a high basis order while at the same time introducing Laplacian regularization.
Finally, much like for EAP-based methods, $q$-space sampling will determine the feasibility and robustness of the estimated MPI indices. Likewise, designing an \textit{ad hoc} sampling would enable additional fine-tuning of MPI performance.

The proposed MPI framework enables us to decouple the simulation of the kernel signal, in particular the chosen simulation parameters, from the estimated quantities (indices) thanks to the propagator formalism.
Independently from how the kernel signal is simulated, or from which microscopic propagator indices are obtained, the proposed MPI framework is general and opens new possibilities to quantitatively describe the diffusion process due to the microstructures without explicit biases due to their mesoscopic directional arrangement within the voxel volume.
Therefore, MPI brings new complementary information to conventional methods that retrieve the Ensemble Average Propagator (EAP).
In fact, while EAP methods provide indices whose values change with the ODF, MPI indices are ODF-invariant and thus relate solely to the microscopic tissue properties such as the volumes, sizes, and shapes of axons and cells (and possibly any other relevant microstructures). 
Consequently, MPI is expected to have a higher \emph{diagnostic} relevance and a more direct interpretability than EAP methods. For example, longitudinal differences in MPI indices in the same subject could be more specifically linked to a change in tissue \emph{integrity} or composition rather than a change in its \emph{orientational organization}. At the same time, MPI provides indices that retain quantitative interpretability of the diffusion process. In fact, much like DTI and MAP-MRI, MPI indices can be used to compare differences with respect to healthy normative values. Thus, further work may address the question of the diagnostic sensitivity of MPI to various pathological conditions of nervous tissue.

\section{Conclusion}
Microscopic Propagator Imaging (MPI) brings about a new diffusion MRI methodology to extract quantitative information about the tissue microstructures while remaining unaffected by their (mesoscopic) orientational organization. This property allows variations in MPI indices to more directly reflect changes in the volumes, sizes, and shapes of axons, cellular somas, and potentially other microstructures. This implies that MPI may offer a higher specificity for these types of microstructural changes compared to other quantitative propagator-based methods, like Diffusion Tensor Imaging and Mean Apparent Propagator MRI, which may translate to a better characterization of the tissue in view of diagnostic and prognostic applications.

\printcredits

\section*{Acknowledgments}
This work received funding from Danmarks Frie Forskningsfond (Independent Research Fund Denmark) with case number 3105-00129B.
The authors thank the University of Verona for the Visting Scholars \& Professors program 2022, which gave the opportunity to initiate the collaboration behind this work.
Data were provided by the Human Connectome Project, WU-Minn Consortium (Principal Investigators: David Van Essen and Kamil Ugurbil; 1U54MH091657) funded by the 16 NIH Institutes and Centers that support the NIH Blueprint for Neuroscience Research; and by the McDonnell Center for Systems Neuroscience at Washington University.

\bibliographystyle{cas-model2-names}

\bibliography{references}


\end{document}